\documentclass[aps,pre,twocolumn,groupedaddress]{revtex4}

\usepackage{longtable}
\usepackage{graphics}
\usepackage{amsmath}
\usepackage{amssymb}
\usepackage{amsfonts}
\usepackage{bm}

\usepackage{epsfig}
\usepackage[american]{babel}
\usepackage[dvipdfm]{color}
\usepackage[latin2]{inputenc}
\usepackage[american]{babel}
\usepackage{color}
\usepackage{graphicx}

\begin{document}

\title{Invaded cluster algorithm for a tricritical point in a diluted Potts model}

\author{I. Balog}
\email{balog@ifs.hr}
\author{K. Uzelac}
\email{katarina@ifs.hr}

\affiliation{Institute of Physics, P.O.Box 304, Bijeni\v{c}ka cesta 46, HR-10001 Zagreb, Croatia}

\begin{abstract}
The invaded cluster approach is extended to 2D Potts model with annealed vacancies by using the random-cluster representation. Geometrical arguments are used to propose the algorithm which converges to the tricritical point in the two-dimensional parameter space spanned by temperature and the chemical potential of vacancies. The tricritical point is identified as a simultaneous onset of the percolation of a Fortuin-Kasteleyn cluster and of a percolation of "geometrical disorder cluster". The location of the tricritical point and the concentration of vacancies for $q=1$, $2$, $3$ are found to be in good agreement with the best known results. Scaling properties of the percolating scaling cluster and related critical exponents are also presented.
\end{abstract}

\pacs{05.50.+q, 64.60.Fr, 75.10.Hk}

\date{\today}

\maketitle

\section{Introduction\label{secintro}}

   Detailed investigations of fractal properties related to criticality were done some time ago for the geometrical phase transitions such as percolation \cite{StaufferStanleyHavlin}. The interest for similar properties in the context of thermal phase transitions \cite{ConiglioKlein,exp2}, less exploited until know, has been renewed recently \cite{JankeSchakel,JankeSchakel1,QianDeng,DengGuoBloete}. Monte Carlo (MC) studies of phase transitions in the last decade have given rise to several cluster algorithms, based on the Fortuin-Kasteleyn (FK) representation of the partition function \cite{FortuinKasteleyn}. In addition to their principal task to reduce the critical slowing down present in the local update algorithms, they have an advantage to offer a better insight into geometrical aspects of phase transitions and represent a natural tool for numerical studies of these phenomena. An algorithm in which this geometrical approach is used in a particular way is the invaded cluster (IC) algorithm, defined by Machta \textit{et al}. \cite{MachtaChayes}. While in the standard cluster approaches, such as the Swendsen-Wang (SW) \cite{SwendsenWang} or Wolff algorithm \cite{Wolff}, the clusters are built for a given temperature, the IC algorithm starts from some geometrical property of the criticality that can be generated by a random process and obtains the critical temperature as an output. It was applied to Ising and Potts models and, later, in a series of other studies, e. g. on the fully frustrated Ising model \cite{frustrIC}, or the XY model \cite{XYIC}. It appears equally efficient in both the second- and first-order phase transitions.

   The tricritical point present in systems which exhibit the changeover from the first- to second-order phase transitions is difficult to access in numerical and finite-size scaling studies due to crossover effects. Even the location of the tricritical point appears to be a difficult task in many cases, from models with long-range interactions \cite{KL00,UzelacGlumac00,ReynalDiep04} to models with the quenched dilution \cite{Mercaldo06}. Standard MC approaches identify this point as the onset of a first-order transition, recognized by two maxima in the free energy distribution which scale as surface \cite{LeeKosterlitz,LandauH} and can be analyzed directly, or from the Binder's fourth cumulant \cite{Binder}. Some alternative numerical approaches were also proposed - e. g. the microcanonical MC study of the tricritical point \cite{Deserno} in Blume-Capel model. In the present work we examine a different approach and extend the IC algorithm by introducing an additional geometrical condition. We present it on the example of a 2D Potts model with the annealed dilution.

The paper is organized as follows: In section \ref{secmodel} we discuss the model and its graphical expansion in the diluted case. In section \ref{secmethod} we present the extension of the IC approach to the tricritical point and explain the algorithm. In section \ref{secresults} we discuss our results for the location of the tricritical point, scaling properties of percolating cluster, and related critical exponents. Section V contains the conclusion.

\section{Model\label{secmodel}}

\indent We consider the $q$ state Potts model \cite{Potts} with annealed vacancies on a square lattice, described by the Hamiltonian
\begin{equation}
\label{dil_Potts}
H=-J\sum_{<i,j>}\big(\delta_{s_i,s_j}t_it_j-1\big)+G\sum_{i}(t_i-1),
\end{equation}
where $s_i$ denotes the $q$-state Potts variable at the site $i$ and the variable $t_i$  takes the values $0$ or $1$ when the site $i$ is empty or occupied, respectively.  The sum is taken over nearest neighbors, $J>0$ is a ferromagnetic coupling, and $G$ is the chemical potential of vacancies. The pure model, where a single thermodynamic parameter -  temperature - is governing a transition, is recovered in the limit $G\rightarrow -\infty$. 

In the pure model the transition is of the second order for $q\le 4$ and of the first order for $q>4$ \cite{Baxter}. The presence of annealed vacancies induces the first-order phase transition, so that for $q<4$ the transition line in the $(T,G)$ parameter space contains both regimes of the first- and second-order phase transition, separated by a tricritical point.

For the pure 2D Potts model, the exact analytical expressions exist both for the critical temperature and critical exponents \cite{Wu}. For a diluted case, exact analytical expressions for tricritical exponents are also available, from the conformal theory \cite{conf1} and Coulomb gas mapping \cite{coulgas1}, while for the location of the tricritical point there are only good approximate results \cite{Deng}.

\indent As shown by Fortuin and Kasteleyn \cite{FortuinKasteleyn}, the pure Potts model is equivalent to the random-cluster model, which may be understood as a generalized percolation model. The random-cluster partition function is given by
\begin{equation}
\label{rand_cluster}
Z=\sum_{\gamma\in \Gamma}p^{b(\gamma)}(1-p)^{E-b(\gamma)}q^{c(\gamma)},
\end{equation}

\begin{equation}
\label{p}
p=1-e^{-\beta J}.
\end{equation}
The summation runs over the set of all the graphs on the lattice $\Gamma$. Each graph represents one choice of placing the bonds on the lattice. The quantity $p$ (where $\beta$ is the Boltzmann factor) may be interpreted as the probability of presence of a bond on an edge, $b(\gamma)$ is the number of bonds in the graph $\gamma$, $E$ is the total number of edges (maximum number of bonds), which for the square lattice of the linear size $L$ is equal to $2L^2$. The entropy factor for each cluster is given by $q$, and $c(\gamma)$ is the number of connected components of a graph. FK clusters have a physical meaning - the probability that two spins separated by a distance $r$ are in the same FK cluster is proportional to the correlation function.

Graph expansion for diluted Potts model has already been studied elsewhere \cite{ChayesMachtaPhysA,KunHu} with a different choice of a Hamiltonian. It is easy to perform a graph expansion for the Hamiltonian (\ref{dil_Potts}) by proceeding along the same lines as in the pure model. By using the equality
\begin{equation}
\label{equality}
e^{\beta J \delta_{s_i,s_j}t_it_j}=[1+(e^{\beta J}-1)\delta_{s_i,s_j}t_it_j],
\end{equation}
%\vskip 20pt
the partition function can be expanded over all the possible graphs on the lattice
%\vskip 20pt
\begin{equation}
\label{dil_Potts_Z0}
Z=e^{E\beta J}\sum_{\gamma\in\Gamma}\sum_{\{s_i\}}\sum_{\{t_i\}}\prod_{i}e^{-\beta G(t_i-1)}\prod_{<i,j>}(e^{\beta J}-1)\delta_{s_i,s_j}t_it_j, 
\end{equation}
where $\{s_i\}$ and $\{t_i\}$ denote the summation over all the configurations of  $s_i $ and  $t_i $ respectively. The structure of graphs remains the same as in the pure model. In each graph $\gamma$ one may separate the summation over the configurations including sites belonging to the clusters of sizes $>1$ from the summation over the "isolated sites" consisting of single spins and vacancies. The summation over isolated sites may be seen as a lattice gas of single spins and vacancies in the field and gives 
%\vskip 20pt
\begin{equation}
\label{latt_gas}
\sum^{n_{is}}_{n_v=0}\binom{n_{is}}{n_v}e^{\beta Gn_v}q^{n_{is}-n_v}=(e^{\beta G}+q)^{n_{is}},
\end{equation}
where  $n_{is}(\gamma)$ is the number of isolated sites on the graph, and $n_v$ is the number of vacancies. The partition function may be written in a more condensed form
%\vskip 20pt
\begin{equation}
\label{dil_Potts_Z1}
Z=\sum_{\gamma\in\Gamma}p^{b(\gamma)}(1-p)^{E-b(\gamma)}q^{\tilde{c}(\gamma)}(e^{\beta G}+q)^{n_{is}(\gamma)},
\end{equation}
to be compared with the pure case Eq. (\ref{rand_cluster}). In the first three factors we recover the same expression as for the pure model except that  $\tilde{c}(\gamma)$ denotes only the number of connected clusters and excludes single spins. In the limit $G  \rightarrow -\infty$ Eq. (\ref{dil_Potts_Z1}) reduces to Eq. (\ref{rand_cluster}) for the pure case. The only change occurs in the contribution to the entropy factor stemming from the isolated sites in each graph. While in the pure case it was equal to $q^{n_{is}}$, in the presence of dilution it corresponds to the one of a lattice gas, given by Eq. (\ref{latt_gas}). The chemical potential $G$, responsible for dilution can be expressed through the parameter 
%\vskip 20pt
\begin{equation}
\label{z}
z=\frac{q}{e^{\beta G}+q},
\end{equation}
which, according to Eq. (\ref{latt_gas}), has the meaning of the {\it a priori} probability to find a single spin on an isolated site on a graph.

\section{Method\label{secmethod}}

In Swendsen-Wang and Wolff algorithms one uses the knowledge of temperature to construct FK clusters by adding bonds with probability $p$. Clusters can then be independently flipped to another Potts state. The opposite happens in the IC algorithm of Machta \textit{et al.} FK clusters are grown by a procedure inspired by the invasion percolation \cite{Wil2}. Bonds are placed at random between neighboring spins in equal states, with probability 1, until a geometrical condition, imposed on FK clusters (e.g. the bond percolation), is achieved. The temperature is then deduced by equating the bond probability $p$ defined in Eq. (\ref{p}) with the ratio of the number of bonds to the number of satisfied neighbors. In particular, if the geometrical condition is the onset of the percolating FK cluster, the output temperature converges to the transition temperature in the thermodynamic limit. In the language of connectivity the appearance of the percolating FK cluster on the lattice corresponds to the state when the correlation length reaches the system size.  

The tricritical point is determined by two parameters, the tricritical temperature $T_t$ and field $G_t$, so the algorithm must include an additional condition besides the percolation of an FK cluster in order to converge to it. In the present 2D case we set the additional condition to be the percolation threshold of a "geometrical disorder cluster" defined as a cluster consisting of vacancies and single spins. That cluster is geometrical in the sense that two adjacent sites containing either a vacancy or a single spin are considered to belong to the same disorder cluster with probability $1$. This choice of the disorder cluster is justified because single spins behave in the same way as vacancies as far as correlations are concerned. It is supported by the picture obtained from the real-space renormalization group where the single spins take part in the renormalization of vacancies \cite{Nienhuis}.

One can also explain the second condition for the tricritical point in terms of the persistence length $\tilde{\xi}$ \cite{Rikvold,Beale}. At the line of second-order transitions the correlation length $\xi$ diverges, but in the case of a tricritical point, there is another length scale in the system, $\tilde{\xi}$, which is related to the size of disorder clusters \cite{Rikvold}. The length $\tilde{\xi}$ is finite on the line of second-order transitions but it diverges at the tricritical point and remains infinite on the line of first-order transitions. Our algorithm generates finite system configurations in which $\xi\propto\tilde{\xi}\propto L$, since the percolation of a geometrical disorder cluster can be interpreted as $\tilde{\xi}$ reaching the size $L$.

\subsection{Algorithm}

An algorithm that is meant for locating the tricritical point should find the percolation of the FK and geometrical disorder clusters at the same time. In this respect we notice that in 2D the simultaneous percolation of two clusters is topologically obstructed unless it occurs in only one direction. Such anisotropic cases occur in a finite lattice and will be facilitated by a diverging persistence length, which occurs when approaching a first-order transition regime. 

Let us outline the basic steps to be explained in detail below. 
We start the MC iterations with random configuration of spins and randomly distributed vacancies with some initial concentration. 
Each MC iteration includes three steps: (i) the formation the FK clusters in the same way as the IC algorithm of Machta \textit{et al.} \cite{MachtaChayes}, (ii) identification of disordered clusters including a check for the percolation, (iii)  randomization of FK clusters and vacancies (keeping the concentration unchanged). The removing or adding of vacancies, necessary to achieve the percolation threshold of the geometrical disorder cluster, is performed in more spaced intervals of MC iterations, and by imposing limitations on the number of vacancies created (destroyed) as described in Sect. \ref{subsecrules}. Namely, when the number of spins does not change, both the mean temperature and the field self-regulate to the transition-line. By changing the number of single spins sufficiently slowly it can be driven along the transition-line to the percolation threshold of the geometrical disorder cluster.   

\subsubsection{Formation of FK clusters}   

The simulation starts with some configuration of the Potts spins and vacancies on the lattice. Bonds are placed randomly, between the Potts spins in equal states. When percolation is achieved, or all the edges have been probed, the procedure is terminated giving a graph consisting of FK clusters. If percolation is achieved, this means that a quasi-critical configuration has been found for a given number of vacancies. 

\indent There are several ways to characterize the onset of a percolating cluster during simulations. Two examples are: (a) winding of a cluster around a lattice with periodic boundary conditions; (b) the span of a cluster becomes equal to the lattice size. The first characterization, termed a topological percolation, will be used in this work - from now on, when percolation on a finite lattice is mentioned it is meant in the sense of (a). 

\indent The percolation of an FK cluster is determined by a procedure which uses the connecting vectors, described in Ref. \cite{MachtaChayes}.

\subsubsection{Identification of geometrical disorder clusters}

Identification of geometrical disorder clusters is similar to the formation of FK clusters. The edges are probed and adjacent isolated sites are connected by imaginary bonds in order to identify all the geometrical disorder clusters. The disorder bonds are said to be imaginary in the sense that they do not have a physical meaning. The formation of geometrical disorder clusters terminates when all the edges have been probed. Percolation is detected in the same way as for the FK clusters.

\subsubsection{Randomization} 

Regardless whether the quasi-tricritical configuration was found or not, after the two cluster formation steps we randomize the configuration and iterate the procedure. The FK clusters in the configuration are randomized by flipping each of them into a new randomly chosen Potts state. The single spins and vacancies are randomized by exchanging their positions with probability $1$. This procedure conserves the number of spins in the system.

\subsubsection{Adding (removing) vacancies and  thermalization\label{subsecrules}}

As far as temperature is concerned, the IC algorithm is self-regulating, as is easily understood in the pure model. If it starts with a configuration corresponding to $T<T_C$, the FK clusters are expected to percolate after a relatively small amount of bonds has been placed, because there is a small number of obstacles for the formation of a percolating cluster. They are caused by the boundaries between the geometrical Potts clusters. The configuration thus generated will correspond to a temperature higher than the starting one. When applied to a configuration corresponding to $T>T_C$, with small clusters, in different Potts states, the IC algorithm meets a lot of obstacles. The number of bonds that need to be placed is quite large, corresponding to a temperature which is lower than the starting one. 

Despite the additional obstacles caused by vacancies, the temperature self-regulation remains in the present algorithm. However, we were not able to find a similarly elegant way to self-regulate the formation of the disorder cluster. Namely, the fact that the concentration of vacancies at the tricritical point is far below the site percolation threshold, indicates that the percolation of geometrical disorder cluster is not random, but strongly correlated. Consequently, the thermalization procedure is important. In order to keep the number of vacancies just at the minimum necessary for the percolation of the geometrical disorder cluster, the system has to be allowed to relax each time that number of vacancies was changed.

To this purpose we introduce the intervals of $\tau$ MC steps during which the concentration of vacancies may vary by limited, very small amount. For each interval $\tau$ we record the number of times $b$ that the geometrical disorder cluster has percolated together with an FK cluster. Further, we set a condition on the fraction $b$ of simultaneous percolation events during the $\tau$ MC steps, to be less than certain small value $b_0$. To obtain the threshold of disorder cluster percolation, the fraction $b_0$ needs to tend to $0$, but for numerical reasons we set it to a small value of the order of $10^{-3}$ to $10^{-2}$.
 
If $b > b_0$ after $\tau$ MC steps, a small, limited number of vacancies is removed and replaced by spins during the next $\tau$ steps. If $b$ is $0$ after $\tau$ MC steps the same limited number of single spins is switched into vacancies during this period. If $b \leq b_0$, the number of spins remains unchanged for the next $\tau$ MC steps. The number of spins that is added or removed has to be small enough not to disturb the effective correlation established between vacancies in geometrical disorder clusters of the tricritical point.

\subsubsection{Geometrical parameters} 

When quasi-tricritical condition is fulfilled, all the relevant data are recorded and used for statistics; if quasi-tricritical configuration has not been found, no record is done.  

   The thermodynamic parameters $T$ and $G$ are calculated from the geometrical quantities. As in the pure case, the temperature is calculated from the bond probability defined in Eq. (\ref{p}), which can be expressed by the ratio  
\begin{equation}
\label{p_alg}
p=\frac{n_b}{n_{ss}},
\end{equation}
where $n_b$ denotes the number of bonds on the lattice and $n_{ss}$ the number of neighboring pairs in the same state. 

The chemical potential $G$ is expressed through the \textit{a priori} probability $z$ defined by Eq. (\ref{z}). It can thus also be expressed as the average fraction of single spins ($n_d$) in the "lattice gas" of isolated sites consisting of vacancies and single spins,

\begin{equation}
\label{z_alg}
z=\frac{n_d}{n_d+n_v}.
\end{equation}

   We calculate the averages $<p>$ and $<z>$, where $<>$ is taken over the ensemble generated by the algorithm. The distribution of variables generated by the algorithm is not canonical, but is sharply peaked around tricritical values since it has a feedback mechanism that adjusts both parameters to the tricritical point. This was already pointed out earlier \cite{MachtaChayes,Moriarty} in the case of IC algorithm for the pure Potts case.

% ===================================================

\section{Results\label{secresults}}

We have considered the dilute Potts model described by the Hamiltonian (\ref{dil_Potts}) in three different cases, $q$ = $1$, $2$, and $3$. 

The calculations included lattices of linear sizes  $24 \leq L \leq 240$ for the cases $q=1$ and $2$, and $24 \leq L \leq 120$ for the case  $q=3$. The statistics used was $3\cdot 10^5$ MC steps for smaller and up to $5\cdot 10^4$ MC steps for larger lattices. The statistics was also reduced to $10^4$ MC steps in the case  $q=3$. 

The intervals $\tau$ described in section III. A. 3. are chosen to be from $5\cdot 10^{2}$ to $10^{3}$ MC steps and the allowed fraction $b_0$ of the order of $10^{-2}$. This value of $b_0$ was found sufficiently small not to produce systematic shift towards the first-order transition which would exceed statistical error bars.

The number of vacancies allowed to be changed after $\tau$ MC steps is chosen to be $L/24$ for $q=1$ and $2$ and $L/12$ for $q=3$. The choice is arbitrary but allows the runs to be made on smaller lattices such as defined above and give reasonable accuracy. Also notice that the ratio of the number of single spins to be changed to the average number of single spins in the system tends to zero as $1/L$, so the non-equilibrium effects decrease as $L$ increases. 

We present first the finite-size results and their extrapolations for the concentration $<t>$ and parameters $p$ and $z$ at the tricritical point. Further, we give the analysis of the scaling properties related to the percolating clusters and evaluate several tricritical exponents. The error bars of Monte Carlo statistics in presented figures are less than symbol sizes.

The results for tricritical exponents are compared to the exact values known from the conformal theory \cite{conf1,conf2} and Coulomb gas map \cite{coulgas1}, while the tricritical values of parameters are compared to very accurate results obtained by the transfer matrix technique \cite{QianDeng}.

\subsection{The concentration}

\begin{figure}[!ht]
\includegraphics{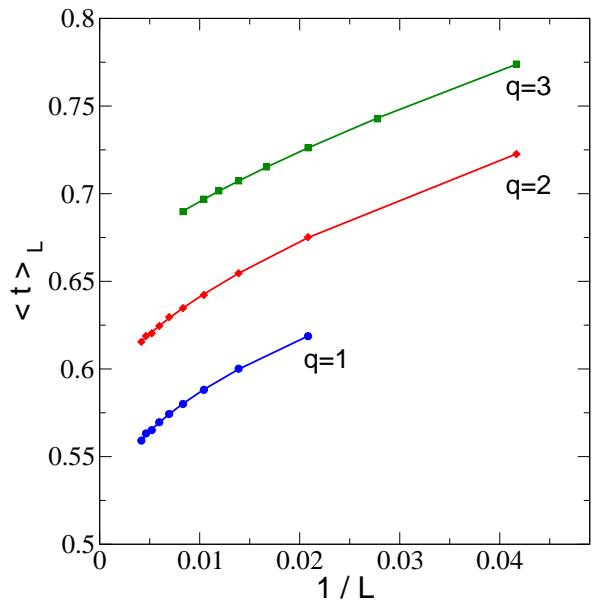}
\caption{(Color online) Plot of  $<t>_L$ vs. $1/L$ for different values of $q$. Plain lines denote the nonlinear fits to the power law form $<t>_L=<t>_t+b\cdot L^{-x}$.}
\label{fig1}
\end{figure}

The quantity for which we have obtained the best precision is the tricritical concentration of spins, $<t>_t$. The finite-size values $<t>_L$ are displayed in Fig. (\ref{fig1}) as a function of the system size $L$.

Simple fits to the power-law form  $<t>_L=<t>_t+b\cdot L^{-x}$ were sufficient to obtain the extrapolations to $L\rightarrow\infty$, which agree to the third digit with the best known results, as shown on Table I.

\begin{table}[!h]
\begin{ruledtabular}
\caption{\footnotesize{Extrapolated values obtained for the tricritical concentration}}
\centering{
\begin{tabular}{ccc}
  $q$ & $<t>$ &$<t>$\footnotemark[1]\\ 
\colrule
  $1$ & $0.499 \pm 0.001$ & $1/2$ \\
  $2$ & $0.579 \pm 0.002$ & $0.57979(1)$\\
  $3$ & $0.652 \pm 0.005$ & $0.65423(4)$\\
\end{tabular}
}\end{ruledtabular}
\footnotetext[1]{Best known results from Ref. \cite{QianDeng}}
\label{tab1}
\end{table}

\subsection{Tricritical parameters}

The temperature and chemical potential are expressed here in terms of the geometrical parameters, the probabilities $p$ and $z$ of the random-cluster expansion ({\ref{dil_Potts_Z1}), and are calculated using Eqs. (\ref{p_alg}) and (\ref{z_alg}).

In Figs. (\ref{fig2}) and (\ref{fig3}) we display all the three sets of finite-size results for  p and z, respectively. They are presented in the form of difference of finite-size results and the expected tricritical values calculated using the parameters given in Ref. {\cite{QianDeng}}. 

The accuracy of any extrapolations is lower by an order of magnitude compared to the one for the particle concentration. The reason for larger discrepancy may be attributed to the fact that these parameters exhibit much larger oscillations than concentration does during simulations and are more susceptible to the thermalization effects. This can be improved by a simple increase of statistics.

\begin{figure}[!hb]
\includegraphics{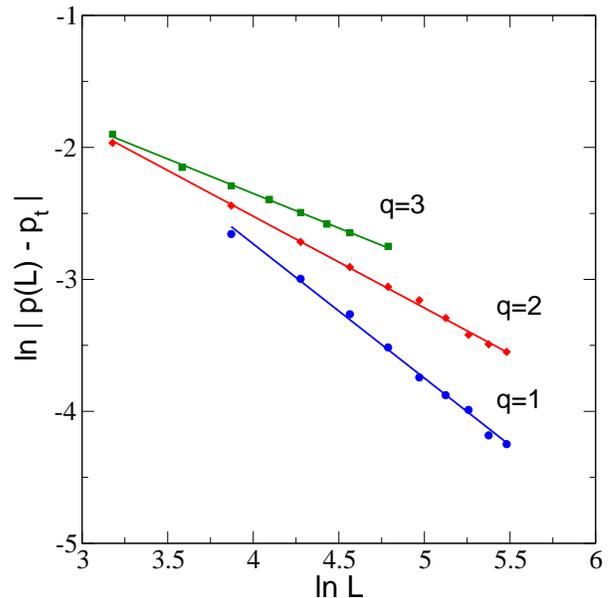}
\caption{(Color online) Double logarithmic plot of $\mid p(L)-p_t\mid$ vs. $L$ for different values of $q$. The best known values for $T_t$ are given in Ref. \cite{QianDeng} and values $p_t$ can be calculated to be $0.8284271$, $0.820168(1)$ and $0.807931(6)$ for $q=1$, $2$, and $3$, respectively.}
\label{fig2}
\end{figure}

\begin{figure}[!ht]
\includegraphics{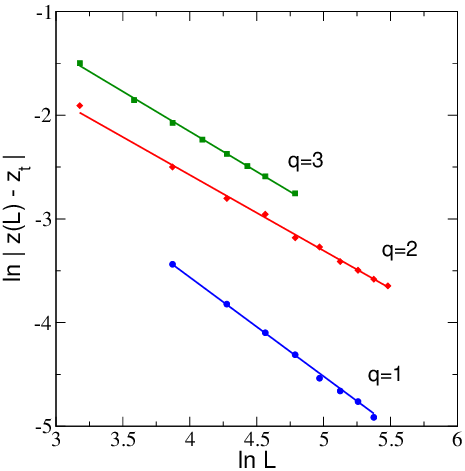}
\caption{(Color online) Double logarithmic plot of $\mid z(L)-z_t\mid$ vs. $L$ for different values of $q$. The best known values for $G_t$ are given in Ref. \cite{QianDeng} and values $z_t$ can be calculated to be $0.02859548$, $0.064220(5)$ and $0.113695(5)$ for $q=1$, $2$, and $3$, respectively.}
\label{fig3}
\end{figure}

As it may be observed in Figs. (\ref{fig2}) and (\ref{fig3}), the convergence of parameters  $p_L$ and $z_L$ can be approximated rather well by a simple power-law form, which can be related to one of the critical exponents.

The results for the convergence exponent  $x_p$, defined as the leading correction for the parameter $p_t$ ($p(L)-p_t \propto L^{-x_p}$)  are presented in Table II for different number of states $q$. They agree well with the values of the subleading critical exponent $y^t_2$.
 
Notice that, in contrast to the ordinary critical point, where the convergence of the critical parameter at criticality is governed by the only relevant critical exponent $y_1$, in the tricritical case there are two relevant exponents which both contribute to finite-size corrections of the critical parameters. In the asymptotic regime the dominant contribution will come from the subleading one and not from $y^t_1$.

\begin{table}[!h]
\begin{ruledtabular}
\caption{\footnotesize{Obtained values for the subleading tricritical exponent $y^t_2$ compared with exact values}}
\centering{
\begin{tabular}{ccc}

 $q$ & $x_p$ & $y^t_2$\footnotemark[1] \\ 
\colrule
 $1$ & $1.02 \pm 0.05$ & $1$ \\
%\hline
 $2$ & $0.69 \pm 0.07$ & $\frac{4}{5}$ \\
%\hline
 $3$ & $0.54 \pm 0.03$ & $\frac{4}{7}$ \\
%\hline
%\hline

\end{tabular}
}\end{ruledtabular}
\footnotetext[1]{Exact results from Ref. \cite{coulgas1}}
\label{tab3}
\end{table}

\subsection{Scaling of FK clusters}

The fractal dimension of the percolating FK cluster is directly related to the anomalous dimension of the order parameter and equal to the magnetic critical exponent $y_h$. Its value at the tricritical point should be different from the one at the second-order transition line, while in the first-order transition regime it is simply equal to the embedding dimension.

In Fig. \ref{fig4} we present the log-log plots of the mass of the percolating FK clusters versus the lattice size. In Table III we present the obtained values for $y_h$ compared to the exact critical and tricritical values. However, in all the three cases considered here, the differences between critical and tricritical values are rather small and of the order of the error bars of our calculations. We can only remark, that the results show a tendency to converge to higher values when only higher sizes are taken into account, which suggests that the considered percolation clusters are indeed tricritical.

More conclusive information, whether the point in the parameter space produced by our algorithm corresponds indeed to the tricritical point and not to a point on the second-order transition line, can be obtained by examining the scaling of the red bonds of the percolating FK cluster.

\begin{figure}[!ht]
\includegraphics{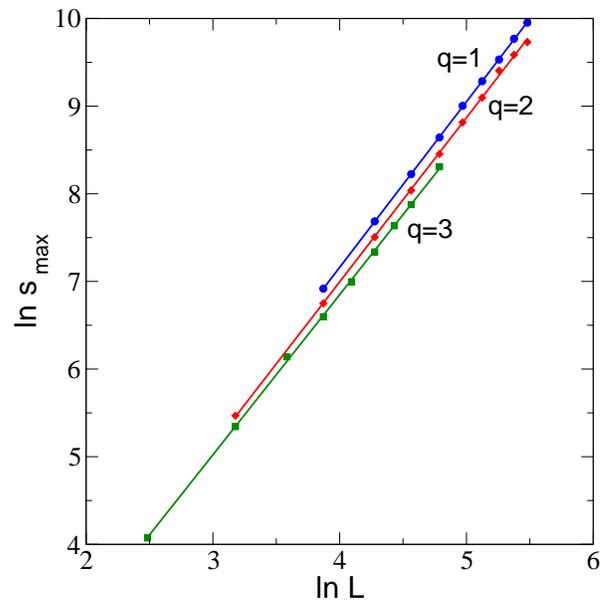}
\caption{(Color online) Double logarithmic plot of percolating FK cluster size $s_{max}(L)$ vs. $L$ for different values of $q$.}
\label{fig4}
\end{figure}

\begin{table}[!h]
\begin{ruledtabular}
\caption{\footnotesize{Obtained values for the magnetic exponent $y^t_h$ compared with exact values for the critical and tricritical cases.}}
\centering{
\begin{tabular}{ccccc}
 $q$ & ${y^t_h}$\footnotemark[1] & ${y^t_h}$\footnotemark[2]& $y_h$\footnotemark[3] & $y^t_h$\footnotemark[3] \\ 
\colrule
 $1$ & $1.89 \pm 0.01$ & $1.89\pm 0.02$ & $\frac{91}{48}$ & $\frac{187}{96}$\\
 $2$ & $1.88 \pm 0.02$ & $1.90\pm 0.02$ & $\frac{15}{8}$ & $\frac{77}{40}$\\
 $3$ & $1.84 \pm 0.03$ & $1.88\pm 0.02$ & $\frac{28}{15}$ & $\frac{80}{42}$\\

\end{tabular}
}\end{ruledtabular}
\footnotetext[1]{Magnetic exponent obtained with all values of $L$}
\footnotetext[2]{Magnetic exponent obtained with $L>72$ for $q=1$ and $2$ and $L>48$ for $q=3$}
\footnotetext[3]{Exact results from Ref. \cite{exp2}}
\label{tab4}
\end{table}

\subsection{Red bonds}

Red bonds are defined as the bonds which, when broken, divide the whole cluster to which they belong into two parts \cite{xx}. While at the second-order transition point the red bonds of the percolating FK cluster have some nontrivial fractal dimensionality, at the tricritical point they scale with a negative exponent, announcing the formation of compact clusters in the neighboring first-order transition regime. The red bonds corresponding to the ordinary critical point of the Hamiltonian (\ref{dil_Potts}) are known to have fractal dimensions of $3/4$, $13/24$, and $7/20$ for q=1,2, and 3 respectively.

Our finite-size results presented in Fig. \ref{fig5}, when fitted to a simple power-law form, give the values cited in Table IV.

\begin{figure}[!ht]
\includegraphics{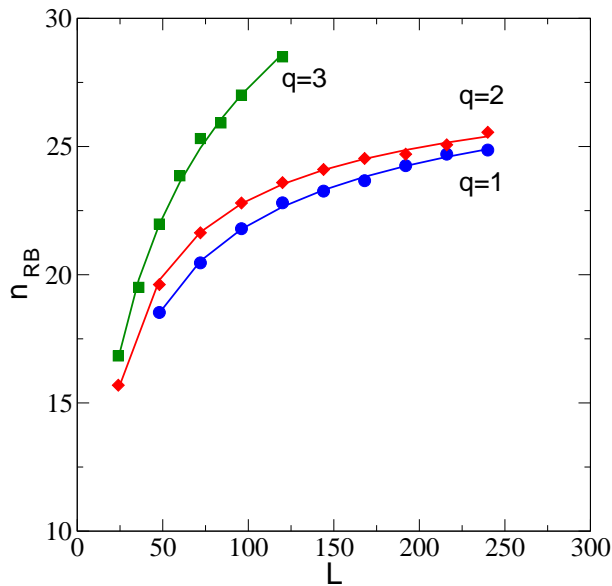}
\caption{(Color online) Number of red bonds on the percolating FK cluster vs. $L$}
\label{fig5}
\end{figure}
\begin{table}[!h]
\begin{ruledtabular}
\caption{\footnotesize{The red bond exponent at the tricritical point compared with exact values}}
\centering{
\begin{tabular}{cccc}
 $q$ & $x_{RB}$\footnotemark[1] & $x_{RB}$\footnotemark[2] &
$x^t_{RB}$\footnotemark[3]\\ 
\colrule
 $1$ & $-0.40\pm 0.06$ & $-0.50\pm 0.08$ & $-\frac{5}{8}$ \\
%\hline
 $2$ & $-0.47\pm 0.04$ & $-0.51\pm 0.07$ & $-\frac{19}{40}$ \\
%\hline
 $3$ & $-0.11\pm 0.07$ & $-0.20\pm 0.08$ & $-\frac{9}{28}$ \\
%\hline
%\hline
\end{tabular}
}\end{ruledtabular}
\footnotetext[1]{Extrapolated form the complete set of values $L$}
\footnotetext[2]{Extrapolated form values $L>72$ for $q=1$ and $2
$ and $L>48$ for $q=3$}
\footnotetext[3]{Exact results Ref. \cite{JankeSchakel}}
\label{tab5}
\end{table}

As it should be expected for relatively small sizes considered in this study,  the finite-size effects are still too strong to allow the calculation of negative exponents with sufficient precision, but the results clearly show negative exponents in all the tree cases.

\section{Conclusion}

We have extended the invaded cluster algorithm to the calculation of the tricritical point.  The algorithm is self-adjusting and locates the position of the tricritical point as the point of a simultaneous onset of the two percolating clusters describing the percolation of order and percolation of disorder due to vacancies. While in the temperature variable the algorithm is completely self-adjusting, like for the simple criticality, in the parameter conjugate to the vacancy concentration the algorithm needs additional fine tuning, in order to insure that the percolation threshold of the disorder cluster is achieved with minimum concentration of vacancies.

The algorithm was illustrated on the example of a dilute Potts model for three different values of $q$. It produced the tricritical concentration with a good precision, and gave reasonably accurate results for the two tricritical parameters. The analysis of the scaling properties of the obtained percolating clusters showed the characteristics proper to the tricritical point. The critical exponents were found in agreement with the tricritical exponents of the considered model.

Let us mention, in the end, that the presented geometrical condition is not the only stopping rule within this invaded cluster procedure that might be constructed for locating the tricritical point. In future, it would be of interest to examine other possibilities (involving e.g. the red bonds), which could be applicable equally to higher dimensions, where the geometrical condition presented here, for topological reasons would not apply.

\end{document}